\documentclass[11pt,twoside]{article}
\usepackage{asp2004}
\usepackage{graphicx}
\usepackage{lscape}
\begin{document}
\title{Observational population studies of cataclysmic variables - The
  golden era of surveys}
\author{B. T. G\"ansicke}
\affil{Department of Physics, University of Warwick, CV4 7AL Coventry,
  UK}

\begin{abstract}
We review the properties of the currently known CV population from an
accountants point of view. In particular, it is examined to what
extent different selection mechanisms (variability, X-ray emission,
colours, spectroscopic properties) affect the observed CV orbital
period distribution. Large-scale surveys provide homogeneously
selected CV samples, for which observational biases can in principle
be modelled. Such samples will eventually provide the essential
observational input for an improved understanding of CV evolution, and
we highlight the impact of several completed and on-going surveys.
\end{abstract}
\thispagestyle{plain}

\section{Introduction}
Historically, cataclysmic variables (CVs) were discovered because of
their \textit{variability}, predominantly being large-amplitude
variables such as dwarf novae or classical novae, e.g. U\,Geminorum
\citep{hind56-1}, but also some irregular low-amplitude systems such
as the polar prototype AM\,Herculis \citep{wolf24-1}. In 20th century,
the rate of discovery of CVs was rather slow until the mid-seventies,
\citet{warner76-1} discussed the properties of some 27 systems with
known orbital periods, and noted the following points: a strong
preference for short-period CVs and a deficiency of CVs in the orbital
period range $1.5-3.25$\,h. It became rapidly clear that the orbital
period distribution of CVs is closely related to the evolution of
these systems, more specifically to the rate at which CVs lose orbital
angular momentum \citep[e.g.][]{eggleton76-1}. To our knowledge, the
first orbital period histogram was published by
\citet{whyte+eggleton80-1}, based on data for 33 CVs, clearly showing
the presence of a ``period gap'' between $2-3$\,h. Both the number of
discovered CVs and the amount of detailed follow-up studies have 
rapidly increased since 1980, stimulating growing activity on the
topic of CV evolution. \textit{Disrupted magnetic braking} became the
standard scenario of CV evolution explaining the presence of the
period gap in the context of an abrupt change in the rate of orbital
angular momentum loss at $P_\mathrm{orb}\simeq3$\,h
(\citealt{rappaportetal83-1, paczynski+sienkiewicz83-1,
spruit+ritter83-1}, see also \citealt{king88-1, howelletal01-1}).

While no globally accepted alternative to the standard model exists so
far, it has been criticised on the base that (a) many of its
predictions are in disagreement with the observations and (b) it is
based on an ad-hoc assumption about the mechanism of orbital angular
momentum loss. The unsettled nature of the field of CV evolution is
reflected in the number of alternatives/modifications to the standard
model that have been published in the recent years
\citep[e.g.][]{clemensetal98-1, mccormick+frank98-1, schenkeretal98-1,
kolb+baraffe99-1, spruit+taam01-1, taam+spruit01-1, kolbetal01-2,
kingetal02-1, king+schenker02-1, schenker+king02-1, schenkeretal02-1,
andronovetal03-1, barker+kolb03-1, taametal03-1, ivanova+taam04-1}.

The purpose of this paper is to review the properties of the known CV
population that currently provides the observational input for CV evolution
theory, and to assess the impact that several large-scale surveys
had/have on this field. 

\begin{figure}
\includegraphics[angle=-90,width=0.5\textwidth]{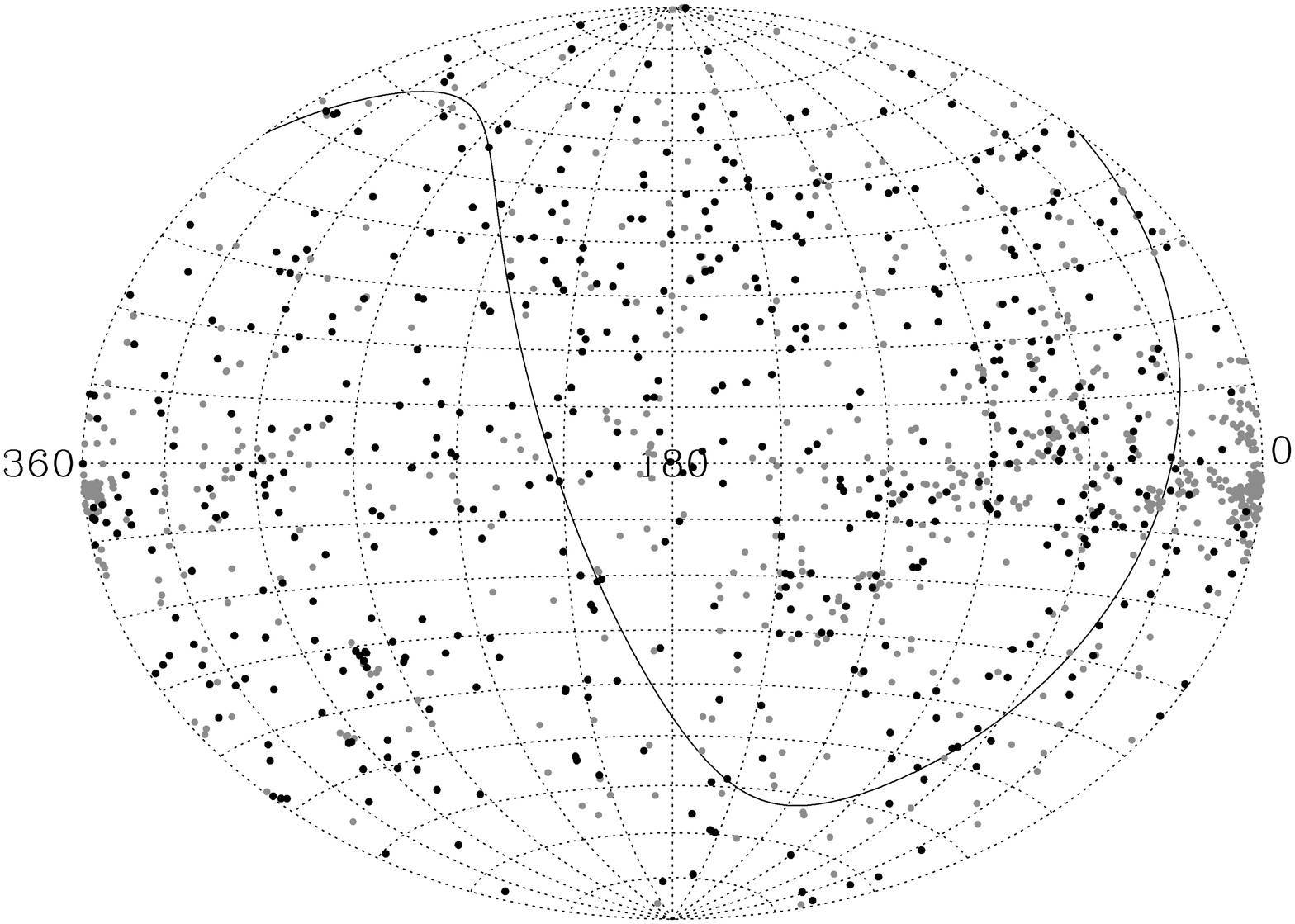}
\includegraphics[angle=-90,width=0.5\textwidth]{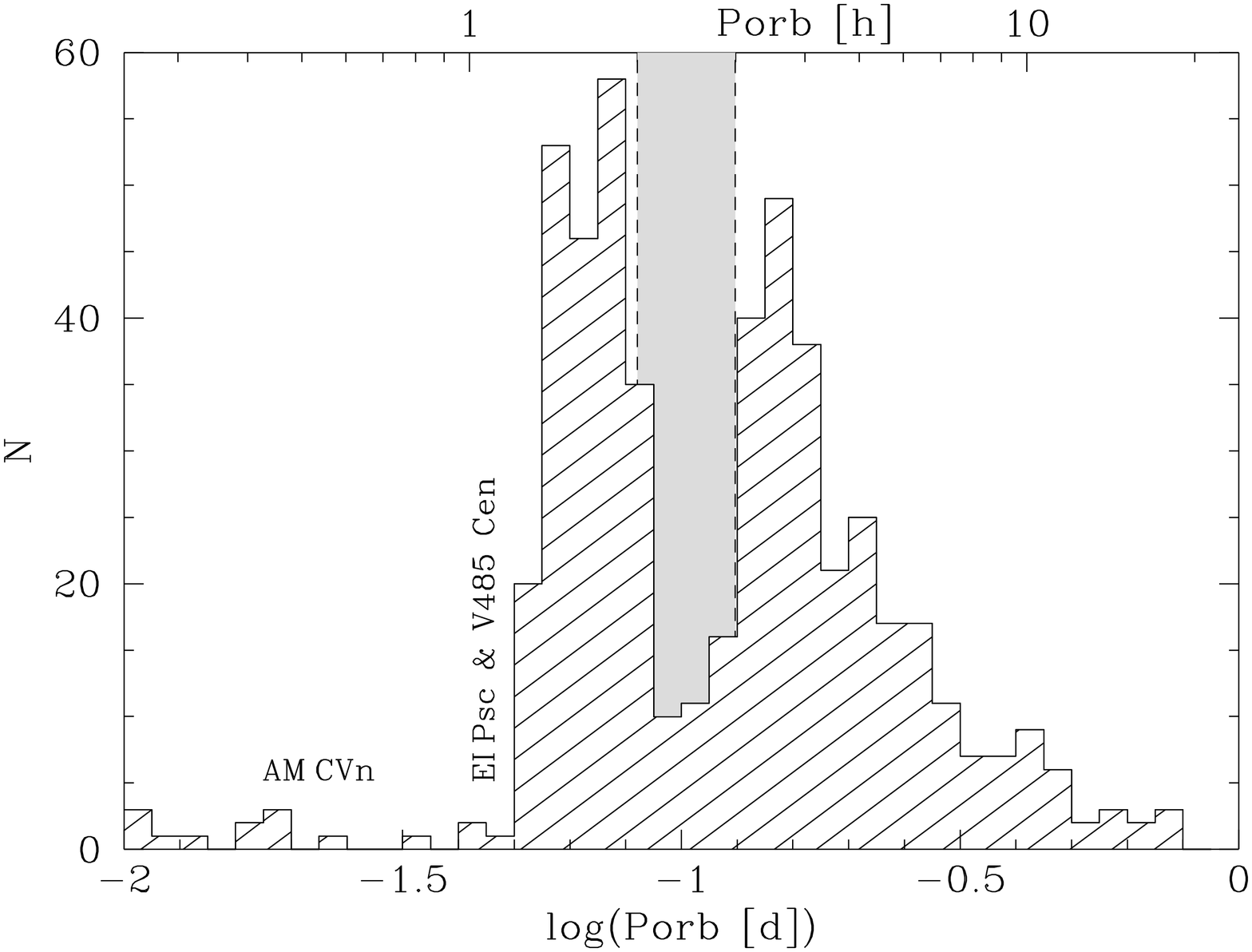}
\caption{\label{f-allcvs}\textit{Left:} The distribution of the known
  CVs from \citet{downesetal01-1} in galactic coordinates. The
  galactic centre wraps around at the right/left edge of the plot,
  the solid line indicates declination $\delta=0$. CVs with a known
  orbital period are shown as black dots, those without as grey dots.
 \textit{Right:} The period distribution of the 531 CVs listed by
 \citet{ritter+kolb03-1}, V7.3. The $2-3$\,h period gap is
 indicated in grey.}
\end{figure}

\section{\label{s-allcvs}The known population of CVs}
The  known population of CVs comprises 531 systems with a
known orbital period \citep[][V7.3]{ritter+kolb03-1}. Qualitatively,
the main features of this period distribution (Fig.\,\ref{f-allcvs})
are a sharp cut-off at $\simeq80$\,min\footnote{Only two CVs with
hydrogen-rich donors are known at shorter orbital periods (EI\,Psc,
$P_\mathrm{orb}=64.2$\,min, \citealt{thorstensenetal02-1}; V485\,Cen,
$P_\mathrm{orb}=59.0$\,min, \citealt{augusteijnetal96-1}), which are
likely to have evolved through a phase of thermal-timescale mass
transfer \citep{gaensickeetal03-1}~--~the other ultrashort-period
systems have degenerate helium-donors (AM\,CVn stars, see Nelemans
this volume).}, a deficit of systems in the gap ($\sim10$\% of
all CV are have orbital periods in the gap), and a steady drop-off in
numbers towards longer orbital periods.

\begin{figure}
\centerline{\includegraphics[angle=-90,width=0.8\textwidth]{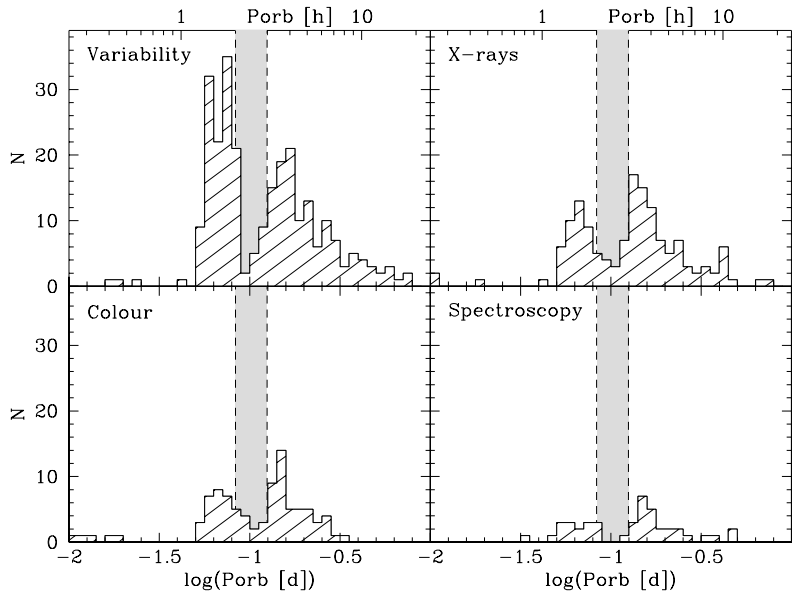}}
\caption{\label{f-porb_discovery} The period distribution of the known
CV population, according to discovery method. }
\end{figure}

The importance of \textit{observational selection} effects became
apparent, at the latest, with the advent of X-ray satellites. Until
the mid-seventies CVs were divided into dwarf novae, novae, and
nova-like variables. The discovery of the variable star AM\,Herculis
being both a strong X-ray source and a strongly magnetic star
\citep{hearnetal76-1, berg+duthie77-1, tapia77-1} added a whole new
aspect to CV research. This illustrates that using the properties of
observed CVs to develop the theory of CV evolution is a risky business
unless the observational biases are fully understood.

Practically all CVs have been discovered using one of the four
following criteria: variability, X-ray emission, peculiar colours
(usually meaning \textit{blue}), and spectroscopic properties
(composite spectra, emission lines). Dividing the known CVs into four
subclasses according to the method of their discovery results in
Fig.\,\ref{f-porb_discovery}. In numbers, 270 CVs were discovered by
variability, 121 by X-ray emission, 91 by colour, and 43 by
spectroscopy. The \textit{variable} CVs are predominantly (86\%)
dwarf/classical novae, the \textit{X-ray} CVs comprise 57\% magnetic
systems, 43\% of the \textit{colour-selected} CVs are novalike
variables, and the spectroscopically selected CVs show no preference
for CV subtype. It is comforting to see that the defining features of
CV evolution, period gap and period minimum, are apparently present in
all four subsamples. Interesting features are the double-peaked
distribution of short-period variable CVs, and the ``spike'' near four
hours in the period distribution of colour-selected CVs. However,
great care has to be taken when interpreting such features, compare
e.g. \citeauthor{hameuryetal90-1}'s (\citeyear{hameuryetal90-1})
prediction with the present-day period distribution of polars with
Fig.\,\ref{f-porb_discovery}. 

\begin{figure}
\includegraphics[angle=-90,width=0.6\textwidth]{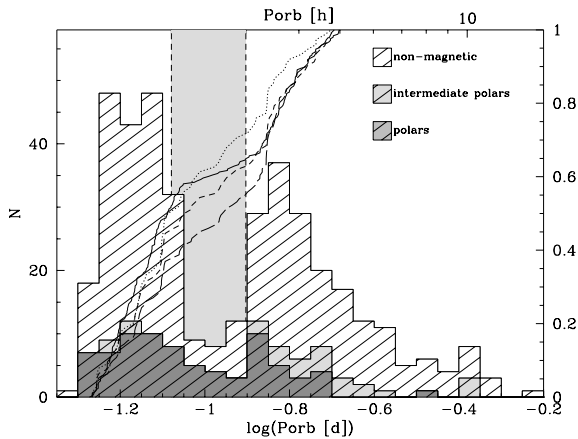}
\hspace*{-0.9cm}
\parbox[t]{0.5\textwidth}{\caption{\label{f-porb_type} Period
distributions divided by strength of the white dwarf magnetic
field. The cumulative distributions are computed between the observed
minimum period (77\,min) and 5\,h. Dotted: polars, short-dashed:
polars and intermediate polars (IPs), long-dashed: polars, IPs and
SW\,Sex stars, solid: dwarf novae and novalike variables excluding
SW\,Sex stars.}}
\end{figure}

Figure\,\ref{f-porb_type} shows the period distribution divided by the
degree of white dwarf magnetism. It has been thoroughly discussed
whether magnetic/non-magnetic CVs evolve in the same way
\citep[e.g.][]{wickramasinghe+wu94-1}, and differences in their
evolution are expected to result in different period
distributions. Judging by eye, one would be prone to suggest that the
period gap is most clearly pronounced for the non-magnetic dwarf
novae/novalike variables, and that the presence of a gap is not
clearly evident for the magnetic CVs. However, comparing the
cumulative distributions of different CV sub-samples based on a
two-sided Kolmogorov-Smirnov test gives a \textit{significantly low}
probability for identical parent CV populations (0.5\%) \textit{only}
if SW\,Sex stars are included in the category of magnetic CVs. All
other reasonable permutations of CV subclasses result in probabilities
for identical parent populations of 16\% to 48\%. It is hence of great
importance to answer the question whether SW\,Sex stars are magnetic
or not \citep[e.g.][]{rodriguez-giletal01-1, hameury+lasota02-1}.

In summary, the interpretation of the overall CV period distribution
and those of different subpopulations in terms of CV evolution is
tempting, but despite the substantial observational efforts invested
over the last three decades, the number of known and well-studied
systems is still too small for a solid statistical assessment.

\section{\label{s-pgcvs}The Palomar-Green survey}

\begin{figure}
\includegraphics[angle=-90,width=0.5\textwidth]{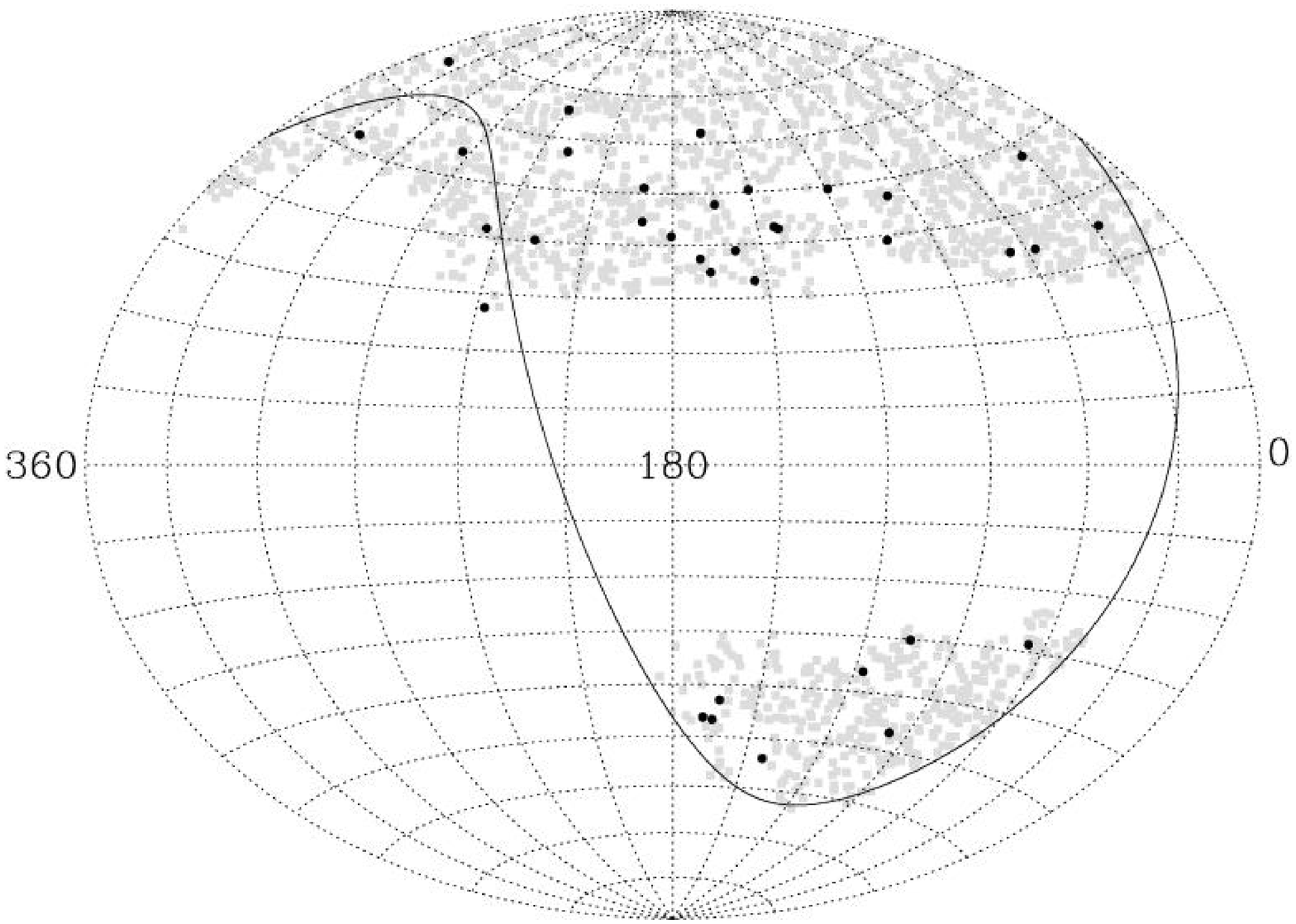}
\includegraphics[angle=-90,width=0.5\textwidth]{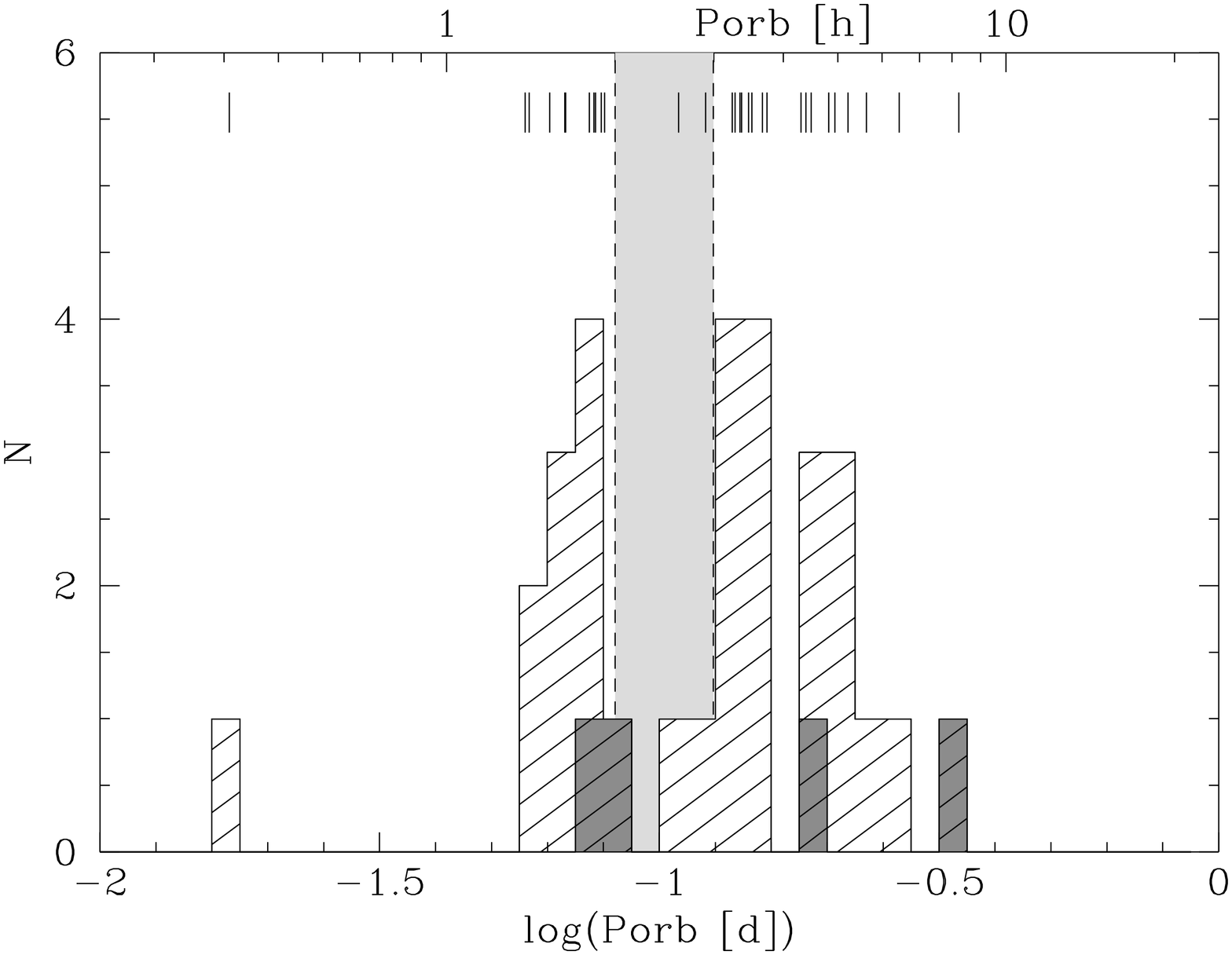}
\caption{\label{f-pg} \textit{Left:} The distribution of CVs (black
dots) from the Palomar Green survey in galactic coordinates. The area
covered by the survey is indicated in grey. \textit{Right:} Orbital
period distribution of the PG CVs. The period gap is indicated in
light grey, previously known systems are shown in dark grey, new
discoveries in white. Tick marks above the histogram indicate the
individual periods of the PG CVs.}
\end{figure}

The photographic $U$ and $B$ Palomar-Green (PG) survey was carried out
using the Palomar 18\arcsec (46\,cm) Schmidt telescope and covered
10\,714\,deg$^2$ at galactic latitudes $|b|>30\deg$ and
declinations $\delta>-10\deg$ with a limiting magnitude $15.5\la
B\la16.7$ \citep{greenetal86-1}. Classification spectroscopy was
obtained for 1878 blue objects with $U-B<-0.46$, resulting in a list
of $\simeq80$ CV candidates. Intensive follow-up studies resulted in
the identification of 31 new CVs \citep[see][and references
therein]{ringwald93-2, ringwald96-1}. In addition to these new
discoveries, the PG CV sample contains 4 previously known CVs. The
distribution of the PG CVs in galactic coordinates, as well as their
orbital period distribution are shown in Fig.\,\ref{f-pg}. Overall,
the PG CV period distribution is similar to that of the entire
observed CV population (Fig.\,\ref{f-allcvs}).

Closer inspection of the PG CVs revealed a group of stars that
\citet{thorstensenetal91-1} coined as \textit{a class of cataclysmic
binaries with mysterious, but consistent, behaviour}: the
SW\,Sextantis stars, with famous founding members being SW\,Sex,
DW\,UMa, PX\,And, and V1315\,Aql. Among the defining characteristics of
these objects are that the vast majority of them have orbital periods
in the range $3-4$\,h, they all have high inclinations\footnote{In
fact, 75\% are eclipsing. This is obviously a selection effect,
implying that a substantial number of CVs with intrinsically similar
properties, but low inclinations have not yet been recognised as
SW\,Sex stars.}, they display single-peaked Balmer and He\,I emission
lines which suffer from transient orbital phase-dependent absorption
in the line cores, moderately strong emission of
He\,II\,$\lambda$\,4686, and the radial velocities measured from the 
emission lines are offset by large degrees with respect to the
expected motion of the white dwarf. All these features are very
difficult to understand within the ``standard'' cartoon of
disc-accreting CVs. Despite intensive efforts, most of the SW\,Sex
behaviour is still mysterious, and the growing number of these systems
(see Sect.\,\ref{s-hqscvs} below) strongly suggests that these objects
represent rather the norm than the exception in the $3-4$\,h period
range. 

Another species of CVs firstly identified in the PG survey are a
number of short-period SU\,UMa dwarf novae that have very short and
coherent super-cycle periods, such as RZ\,LMi, ER\,UMa and V1159\,Ori
\citep{robertsonetal95-1}. The frequent outbursts suggest that these
systems have, despite their short orbital periods, rather high mass
transfer rates.

\section{\label{s-rosatcvs}CVs discovered by ROSAT}

\begin{figure}
\includegraphics[angle=-90,width=0.5\textwidth]{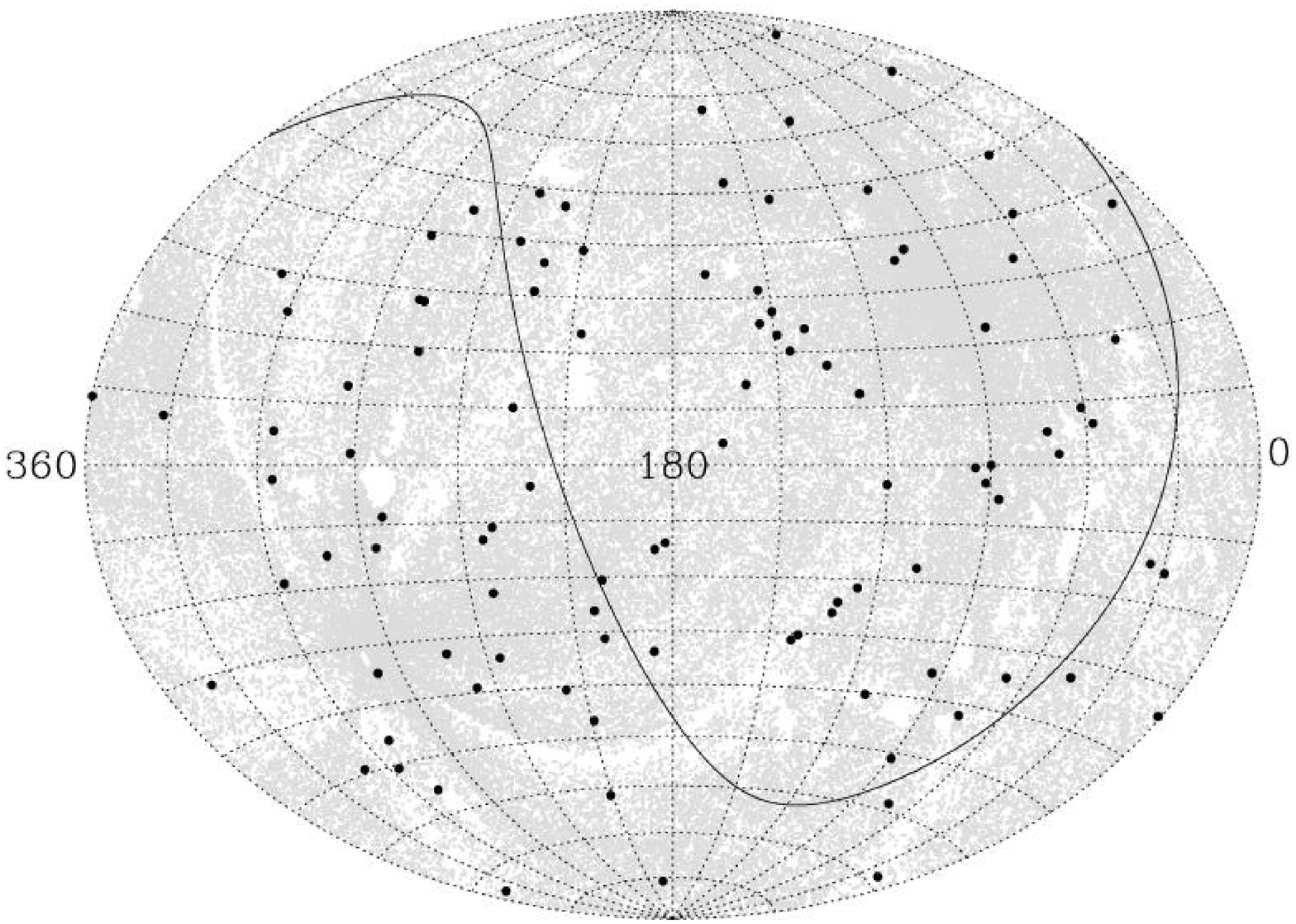}
\includegraphics[angle=-90,width=0.5\textwidth]{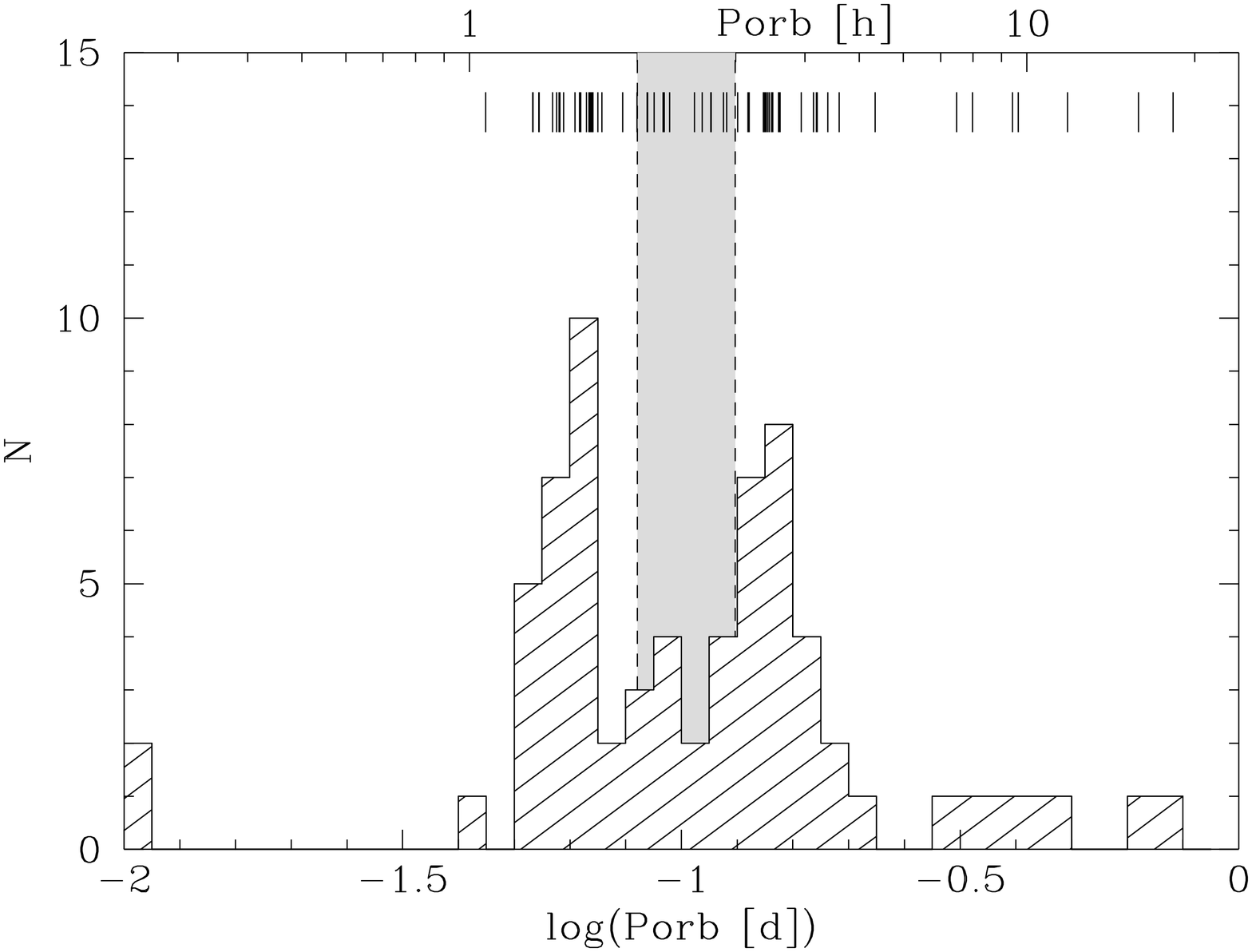}
\caption{\label{f-rosatcvs} \textit{Left:} Galactic distribution of
    CVs discovered with ROSAT. \textit{Right:} Orbital period
    distribution of the ROSAT CVs. See Fig.\,\ref{f-pg} for details.}
\end{figure}

The ROSAT/PSPC all sky survey was the first imaging survey in soft
X-rays ($0.1-2.4$\,keV), and resulted in the identification of
105\,924 X-ray sources \citep{vogesetal99-1,
vogesetal00-1}. Additional X-ray sources were detected in the WFC
survey \citep{poundsetal93-1, pyeetal95-1}, as well as in pointed PSPC
and HRI observations. As already outlined  in Sect.\,\ref{s-allcvs},
polars and intermediate polars (IPs) are the most prominent X-ray
emitters amongst CVs, and consequently the ROSAT mission has had a
huge impact on the statistics of magnetic CVs, doubling the number of
known systems (\citealt{haberl+motch95-1, beuermannetal99-2,
thomasetal98-1}). However, also a substantial number of non-magnetic
CVs were identified during follow-up studies of ROSAT sources, such as
the optically very bright ($V\simeq12.6$) CV RX\,J1643.7+3402
\citep{mickaelianetal02-1}, with the total of ROSAT-lead CV
discoveries exceeding 100. Figure\,\ref{f-rosatcvs} shows the orbital
period distribution of 69 ROSAT CVs for which follow-up studies have
been carried out.

Among the potentially most important discoveries made by ROSAT are the
two ultra-short period systems RX\,J0806.3+1527
($P_\mathrm{orb}=321$\,s, \citealt{burwitz+reinsch01-1,
ramsayetal02-1}) and V407\,Vul\,=\,RX\,J1914.4+2456
($P_\mathrm{orb}=569$\,s, \citealt{haberl+motch95-1,
cropperetal98-2}), which are thought to be double-degenerate CVs. If
these periods should be confirmed to be the orbital periods, these
systems would be the shortest period binaries known, and represent
excellent candidates for the detection of gravitational waves.

\section{\label{s-eccvs}The Edinburgh-Cape survey}

\begin{figure}
\includegraphics[angle=-90,width=0.5\textwidth]{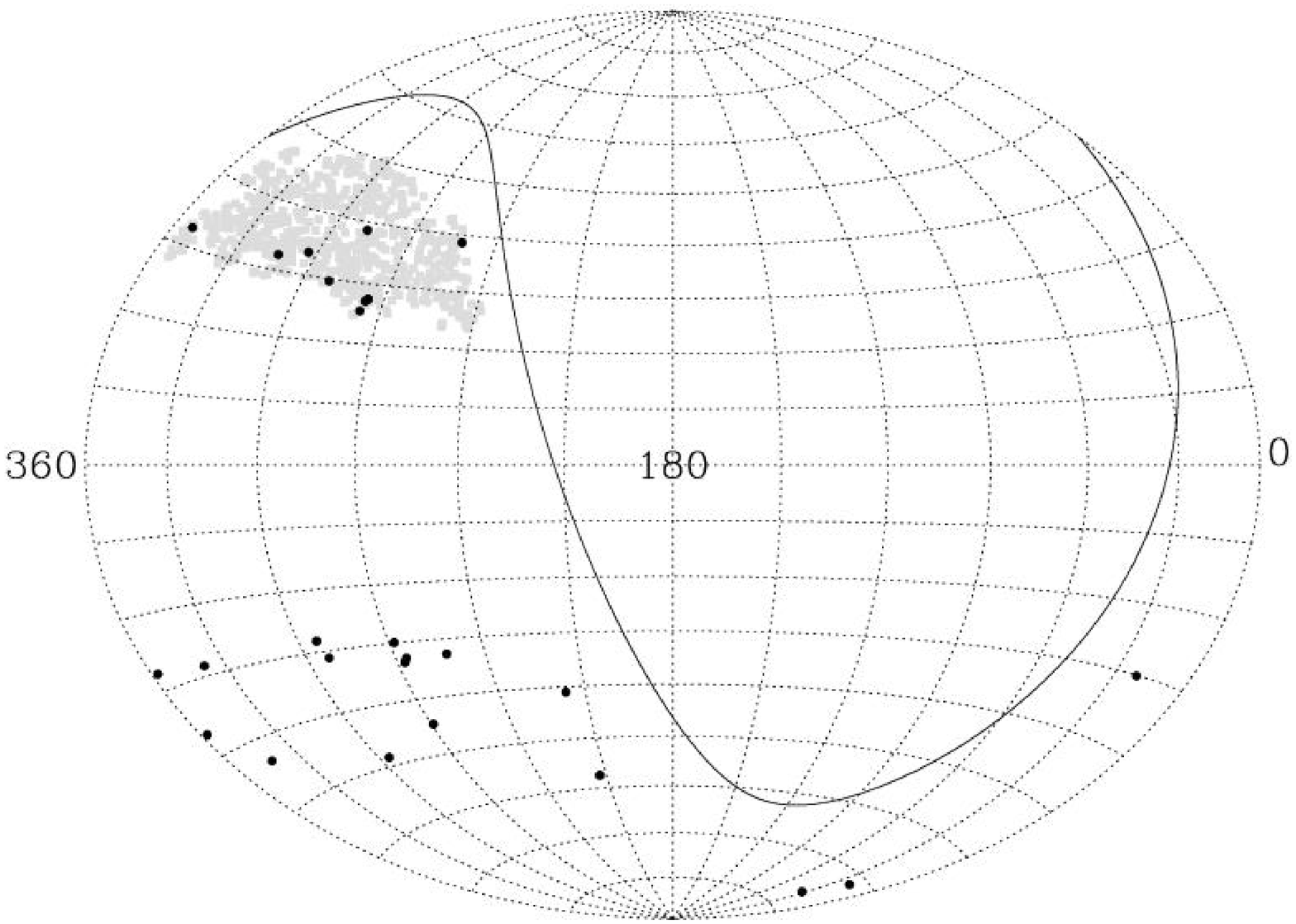}
\includegraphics[angle=-90,width=0.5\textwidth]{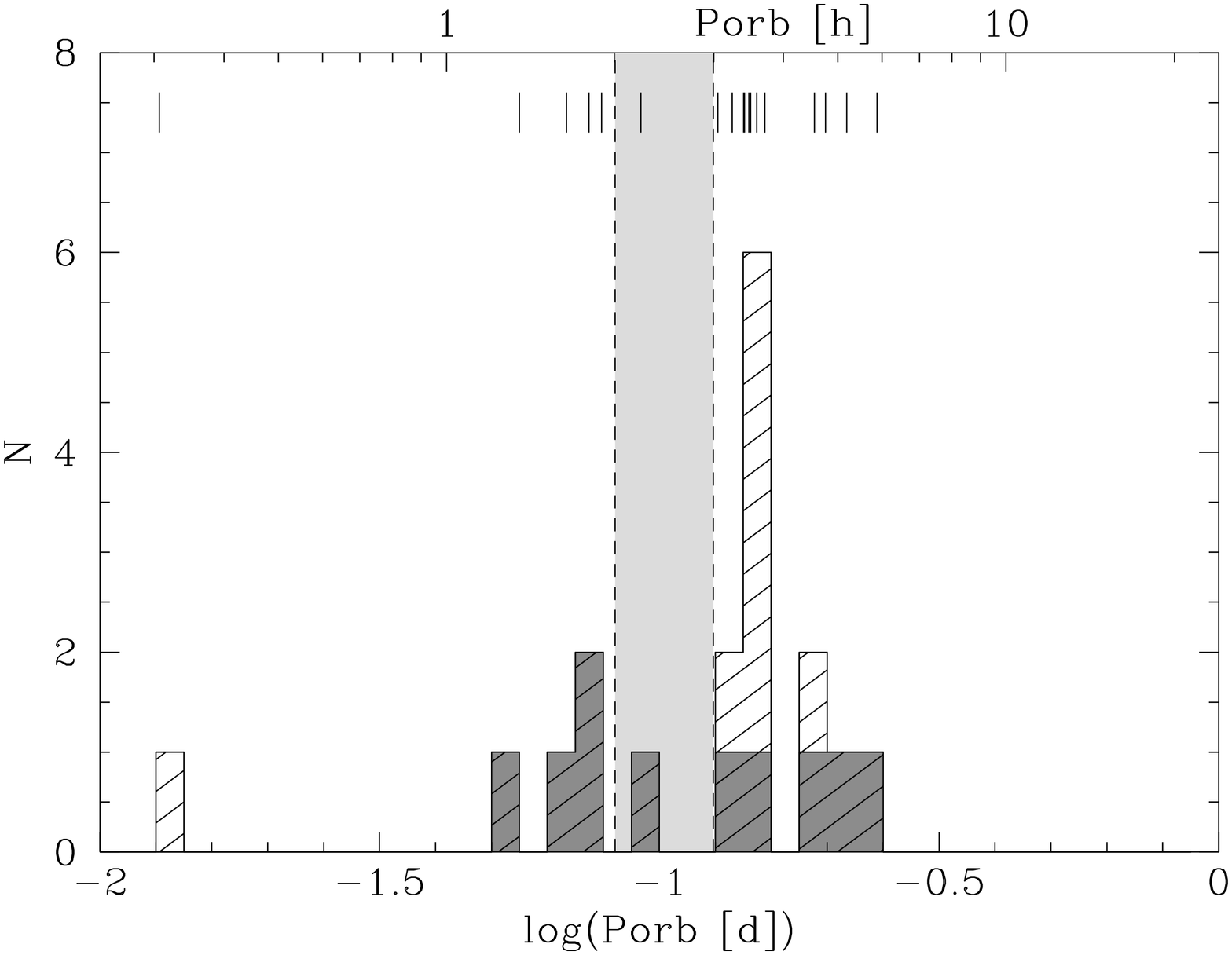}
\caption{\label{f-eccvs} \textit{Left:} Galactic distribution of CVs
from the Edinburgh Cape survey. \textit{Right:} Orbital period
distribution of the EC CVs. See Fig.\,\ref{f-pg} for details. The
total coverage of the EC survey (grey dots) is not known in detail, as only one
EC Zone has been published so far \citep{kilkennyetal97-1}.}
\end{figure}

The Edinburgh-Cape survey has been very similar to the Palomar-Green
survey (Sect.\,\ref{s-pgcvs}) in scope and specifications. The imaging
survey was carried out using the 1.2\,m UK Schmidt telescope at
the AAO, obtaining $U$ and $B$ photographic plates with a limiting
magnitude of $B\simeq18$ and covering $|b|>30\deg$,
$\delta<0\deg$; spectroscopic and photometric follow-up
observations have been obtained at SAAO for objects with $U-B<-0.4$
and $B\la16.5$ \citep{stobieetal97-1, kilkennyetal97-1}. A total of 27
CVs are contained in  the EC survey, of which 15 new discoveries
\citep{chenetal01-1}. Period determinations exist so far for about
half of the new systems (Fig.\,\ref{f-eccvs}). 

It is intriguing that most of the EC CVs with a measured orbital
period are found near $\sim3-4$\,h, i.e. the range where the PG survey
established SW\,Sextantis stars as a homogeneous class of objects
(Sect.\,\ref{s-pgcvs}). It is not yet clear how many of the EC CVs may
share the defining properties of the SW\,Sex stars.

An object of particular interest that has emerged from the EC survey
is the eclipsing detached white dwarf/red dwarf binary
EC\,13471--1258. \citet{odonoghueetal03-1} presented a very detailed
follow-up analysis of this object, which suggests that its donor star
is very close to being Roche-lobe filling, and that the white dwarf is
rotating fairly rapidly. Combining both pieces of evidence,
\citet{odonoghueetal03-1} suggested that EC\,13471--1258 may be a
hibernating CV. Confirming this hypothesis would be
extremely important in the context of \citeauthor{sharaetal86-1}'s
(\citeyear{sharaetal86-1}) nova hibernation scenario.

\section{\label{s-hqscvs}The Hamburg Quasar Survey}

\begin{figure}
\includegraphics[angle=-90,width=0.5\textwidth]{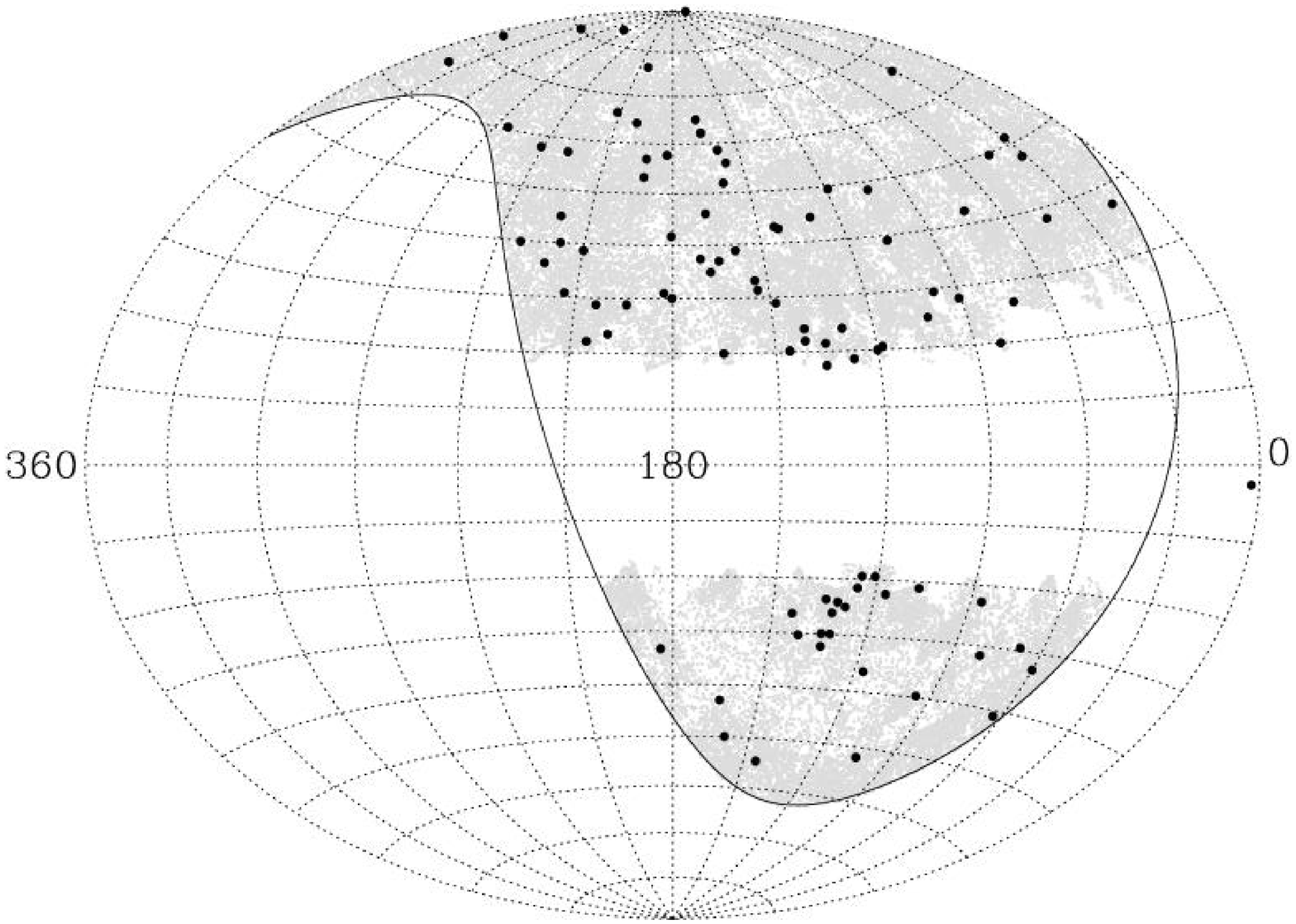}
\includegraphics[angle=-90,width=0.5\textwidth]{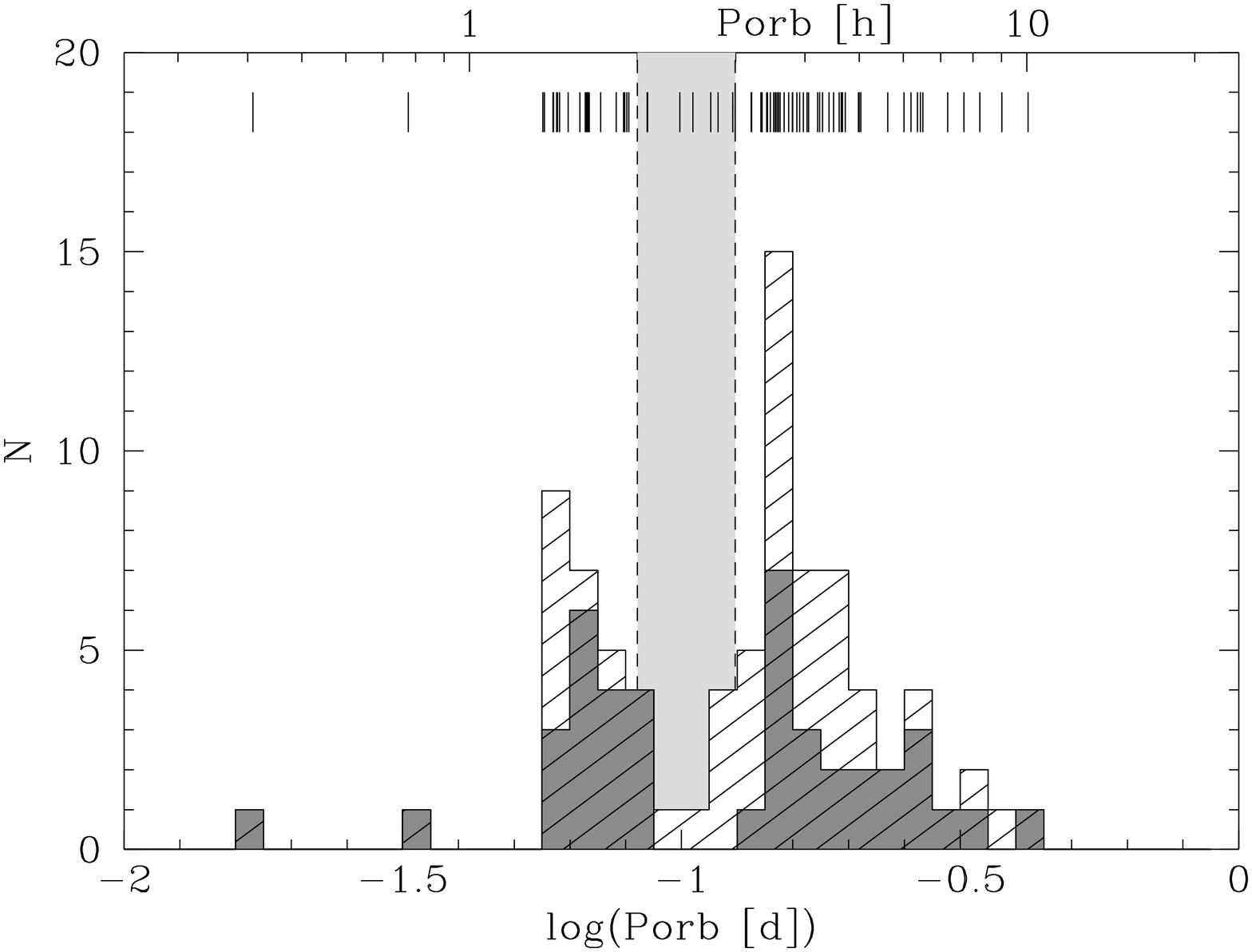}
\caption{\label{f-hqscvs}\textit{Left:} Galactic distribution of CVs
    from the Hamburg Quasar Survey. \textit{Right:} Orbital period
    distribution of the HQS\,CVs. See Fig.\,\ref{f-pg} for details.}
\end{figure}

The Hamburg Quasar Survey (HQS) is an objective prism Schmidt survey
carried out with the 80\,cm Schmidt telescope at Calar Alto, covering
$\simeq13\,600\,\mathrm{deg}^2$ at $|b|>20\deg$ and $\delta>0\deg$ with a
limiting magnitude of $17.5\la B\la18.5$ \citep{hagenetal95-1}. The
spectral range covered by the photographic plates is
$\sim3400-5400$\,\AA, and all fields were observed at least twice to
allow the investigation of variability. \citet{gaensickeetal02-2}
selected CV candidates from the HQS based on (1) the detection of
Balmer line emission and (2) blue colour plus variability.
Identification spectroscopy of these candidates, primarily carried out
using the Calar Alto 2.2\,m telescope, led to the identification of 53
new CVs. \citet{gaensickeetal02-2} carefully investigated the CV
detection efficiency of the HQS as well as possible selection effects
using 84 previously known CVs contained in the survey, and found that
the applied selection criteria recover $\simeq90$\% of the known
short-period ($P_\mathrm{orb}<2$\,h) CVs. This fraction drops to
$\simeq40$\% above the period gap, as many disc-dominated long-period
CVs have weak or no Balmer emission. A conservative estimate of the
completeness is that the HQS allows to identify all CVs with
$B\la16.5$ and an $\mathrm{H}_\beta$ equivalent width $\ga10$\,\AA. So far the
periods of 40 new HQS CVs have been measured (Fig.\,\ref{f-hqscvs}),
follow-up studies of the remaining 13 systems are underway. Whereas the
period distribution of the HQS CVs is still somewhat preliminary, with
25\% of the new CVs lacking a $P_\mathrm{orb}$ measurement, it is
worth to note the following points.

(1) Only relatively few new CVs have been found below the gap, despite
the fact that \citet{gaensickeetal02-2} convincingly demonstrated that
the applied selection criteria are most efficient for short-period
CVs. This clearly implies that most of the short-period CVs in the
parameter space covered by the HQS (sky area, magnitude range, Balmer
line equivalent width) \textit{have been previously identified},
either because of their outbursts (dwarf novae) or their X-ray
emission (polars). In fact, \textit{most} of the new dwarf novae
discovered in the HQS have rather long outburst intervals, e.g. the
SU\,UMa systems KV\,Dra\,=\,HS\,1449+6415 \citep{nogamietal00-1},
HS\,2219+1824 \citep{rodriguez-giletal04-3}, or the deeply eclipsing
$P_\mathrm{orb}=4.2$\,h dwarf nova GY\,Cnc\,=\,HS\,0907 +1902
\citep{gaensickeetal00-2}. The shortest-period CV from the HQS,
HS\,2331 +3905, had no recorded outburst so far, resembling in many
other aspects the (in)famous WZ\,Sge \citep{araujo-betancoretal04-2}.

(2) A very large number of CVs is found with orbital periods in the
range $3-4$\,h (Aungwerojwit et al. this volume; Rodr\'iguez-Gil this
volume), of which a large fraction are SW\,Sextantis stars. This
includes four previously known systems (PX\,And, BH\,Lyn, WX\,Ari, and
LX\,Ser), and 7 new discoveries, HS\,0357 +0614\,=\,KUV\,03580+0614
\citep{szkodyetal01-1}, HS\,0455+8315 \citep{gaensickeetal02-3},
HS\,0551+7241 \citep{dobrzyckaetal98-1}, HS\,0728+6738
\citep{rodriguez-giletal04-2}, HS\,0129+2933, HS\,0220+0603, and
HS\,1813+6122 (Rodr\'iguez-Gil, this volume). The large number of new
SW\,Sex stars found in the HQS (25\% of the whole class) is compatible
with the properties of these objects: while they have relatively
strong emission lines, they are weak X-ray emitters, and exhibit only
moderate variability, making them inconspicuous for the two most
widely used discovery mechanisms. Whereas the PG survey already
suggested that SW\,Sex stars may be fairly common, our follow-up work
on HQS CVs clearly demonstrates that SW\,Sex stars are the dominant species
of systems in the $3-4$\,h period range. The nature of the SW\,Sex
stars is still intensively debated \citep[e.g.][]{hoardetal03-1}. A
number of arguments indicate that these systems have mass transfer
rates well in excess of those predicted by angular momentum loss
through magnetic braking (e.g. the very hot white dwarf primaries
found in SW\,Sex stars, \citealt{araujo-betancoretal03-1}), and it has
been proposed that they may contain weakly magnetic white dwarfs
\citep{rodriguez-giletal01-1, hameury+lasota02-1}. Investigating in
detail the properties of \textit{all} CVs in the $3-4$\,h period range
appears now to be a key project for our overall understanding of CV
evolution.

The HQS has turned out to be efficient in adding new members to the
small class of IPs, too, which also do not display large-amplitude
variability. Specifically, we have identified HS\,0618+7336\,=\,1RXS
J062518.2+733433 as a rather ``normal'' IP ($P_\mathrm{orb}=283$\,min,
$P_\mathrm{spin}=19.8$\,min: \citealt{araujo-betancoretal03-2}),
HS\,0756+1624\,=\,DW\,Cnc as one of the few short-period IPs, very
similar to V1025\,Cen ($P_\mathrm{orb}=86.1$\,min,
$P_\mathrm{spin}=38.6$\,min: \citealt{rodriguez-giletal04-1}), and an
unusual long-period IP ($P_\mathrm{orb}=265$\,min,
$P_\mathrm{spin}=69.2$\,min: Rodr\'iguez-Gil et al. in prep).

As with the other surveys mentioned so far, also the HQS has unearthed
a peculiar group of systems, the magnetic white dwarf/red dwarf
binaries HS\,1023 +3900 =\,WX\,LMi \citep{reimersetal99-1} and
HS\,0922+1333 \citep{reimers+hagen00-1}. Spectroscopically, these
systems are characterised by a very cold white dwarf, a late-type
stellar companion, and extremely strong and sharp cyclotron emission
lines arising from an accretion shock close to the white dwarf
surface. Contrary to all other known polars, these systems have never
been found in a state of high mass transfer, and were consequently
named \textit{low accretion rate polars}. Additional systems of this
kind were identified in the SDSS (Sect.\,\ref{s-sdsscvs}), and it may
well be that these objects are not CVs, but \textit{pre-polars}, with
the white dwarf accreting from the wind of the donor star (see Schmidt
this volume; Webbing this volume).

\begin{figure}
\includegraphics[angle=-90,width=0.5\textwidth]{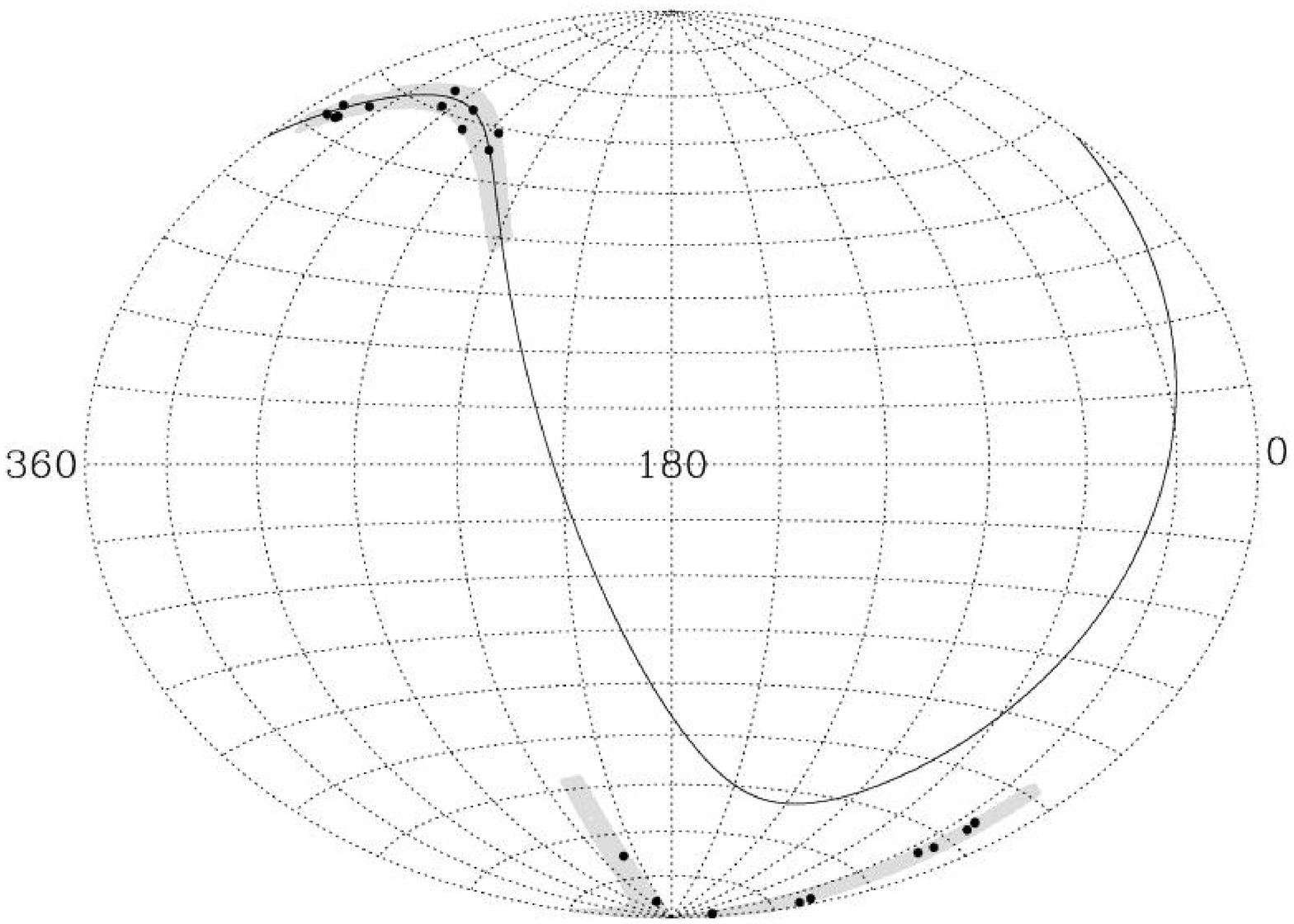}
\includegraphics[angle=-90,width=0.5\textwidth]{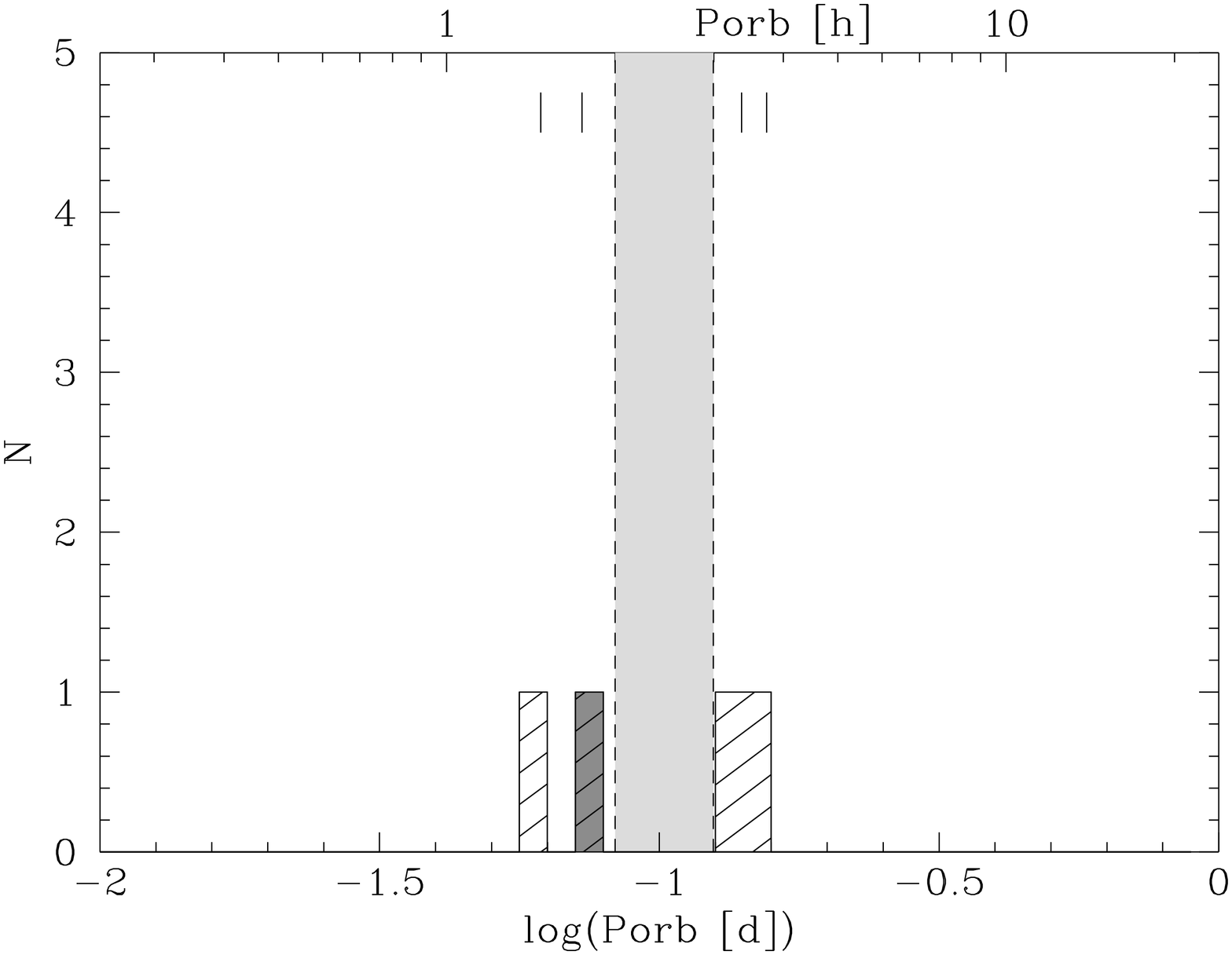}
\caption{\label{f-2qzcvs} \textit{Left:} Galactic distribution of
CVs from the 2dF Quasar Survey
survey. \textit{Right:} Orbital period distribution of the 2QZ
CVs. See Fig.\,\ref{f-pg} for details.}
\end{figure}

Finally, two extremely interesting objects have recently been
discovered in the HQS. The detached white dwarf/red dwarf binary
HS\,2237+8154 resembles EC\,13471--1258 discussed in
Sect.\,\ref{s-eccvs} in that its donor star is almost Roche-lobe
filling. In addition, it contains a very cold/old white dwarf, and its
orbital period is 178\,min, just at the upper edge of the period
gap. \citet{gaensickeetal04-1} suggest three alternative evolutionary
states for the system, being a pre-CV shortly before the start of mass
transfer, a CV that has stopped mass transfer and is entering the
period gap, or, in analogy to EC\,13471--1258, a hibernating
nova. HS\,2331+3905 is probably the most complex CV found to date
\citep{araujo-betancoretal04-2}. It contains a cold white dwarf and
most likely a brown dwarf donor, and shows all signs of an extremely
low mass transfer rate. The system is eclipsing with a period of
81\,min, but also displays a photometric period of 83.4\,min which we
interpret as a permanent superhump. HS\,2331+3905 is the brightest CV
where a pulsating white dwarf, which is also very likely a rapid
rotator, was found. Most exotically, however, is the presence of a
large-amplitude radial velocity variation in the emission line wings
with a period of 3.5\,h, which is in no way commensurate with the
orbital period, and which may be related to a precessing warped inner
disc.

\section{\label{s-2qzcvs}The 2dF Quasar Survey}

The 2dF Quasar Survey (2QZ) is a spectroscopic survey of
colour-selected quasar candidates with $18.4\la B\la20.9$ using the
AAT/2dF and covering $740\,\mathrm{deg}^2$. The candidates were
selected by $U-B\la0.36$ on UKST photographic plates, cutting out the
main sequence \citep{boyleetal00-1}. The total spectroscopic data base
comprises some 48\,000 spectra. 21 new CVs were identified so far, and
\citet{marshetal02-1} reported first follow-up studies (see
Fig.\,\ref{f-2qzcvs}). The 2QZ CVs represent the faintest CV sample
obtained so far.

\section{\label{s-sdsscvs}The Sloan Digital Sky Survey}

The Sloan Digital Sky Survey (SDSS) is by far the largest of the
surveys described in this paper. It is carried out using a
purpose-built 2.5\,m telescope in New Mexico, and has been designed to
obtain deep ($g\la23$) imaging in five broad-band filters and
spectroscopy of $\sim10^6$ colour-selected objects with a limiting
magnitude of $g\simeq20$ at $|b|>30\deg$ \citep{yorketal00-1}. The
main purpose of the SDSS is to obtain an enormous photometric and
spectroscopic data base of galaxies and quasars. QSO candidates for
the spectroscopy are selected by excluding main sequence stars, white
dwarfs, and white dwarf/red dwarf binaries in the four-dimensional
($u-g$, $g-r$, $r-i$, $i-z$) colour space
\citep{richardsetal02-1}. For the discovery of CVs, this broad
selection represents an enormous progress over the earlier
``blue-only" surveys, such as the PG survey (Sect.\,\ref{s-pgcvs}),
and the SDSS CVs will represent the most unbiased sample obtained so
far. At the time of writing, the third data release (DR3) has been
made public, which provides imaging data for $5282\,\mathrm{deg}^2$
and spectroscopy for 528\,640 objects \citep{abazajianetal04-2}. SDSS
up to DR3 has resulted in the spectroscopic identification of 131 CVs,
of which 113 are new discoveries (\citealt{szkodyetal02-2,
szkodyetal03-2, szkodyetal04-1}, Szkody priv. comm.)\footnote{Two
further SDSS data releases are planned, and the total number of SDSS
CVs is expected to be $\simeq200$. Alas, by the time that the funding
for SDSS ends (July 2005), the covered survey area will substantially
fall short of the planned $10\,000\,\mathrm{deg}^2$.}.  The broad
colour range covered by SDSS is reflected in the very broad variety of
subtypes among the SDSS CVs.

\begin{figure}
\includegraphics[angle=-90,width=0.5\textwidth]{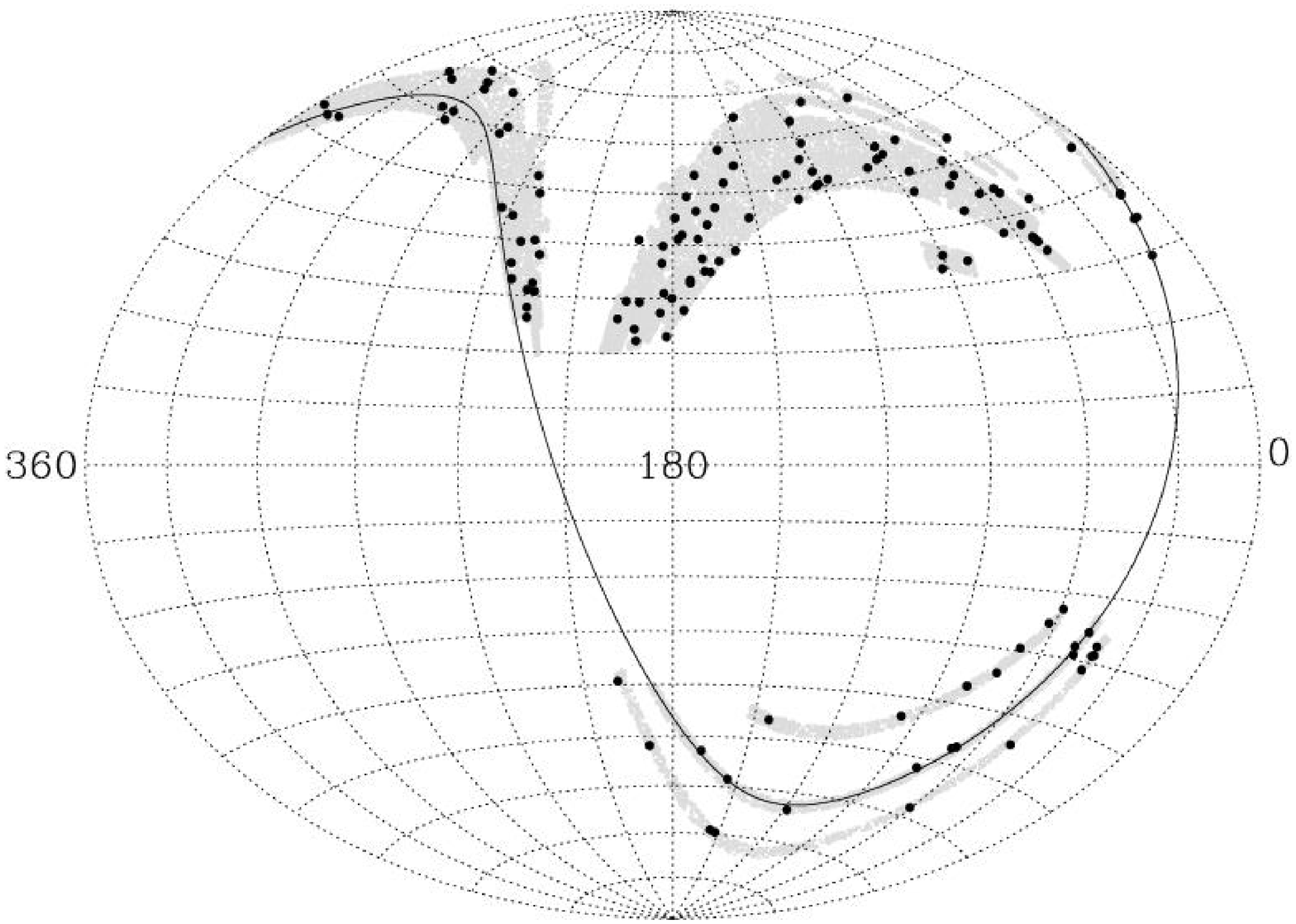}
\includegraphics[angle=-90,width=0.5\textwidth]{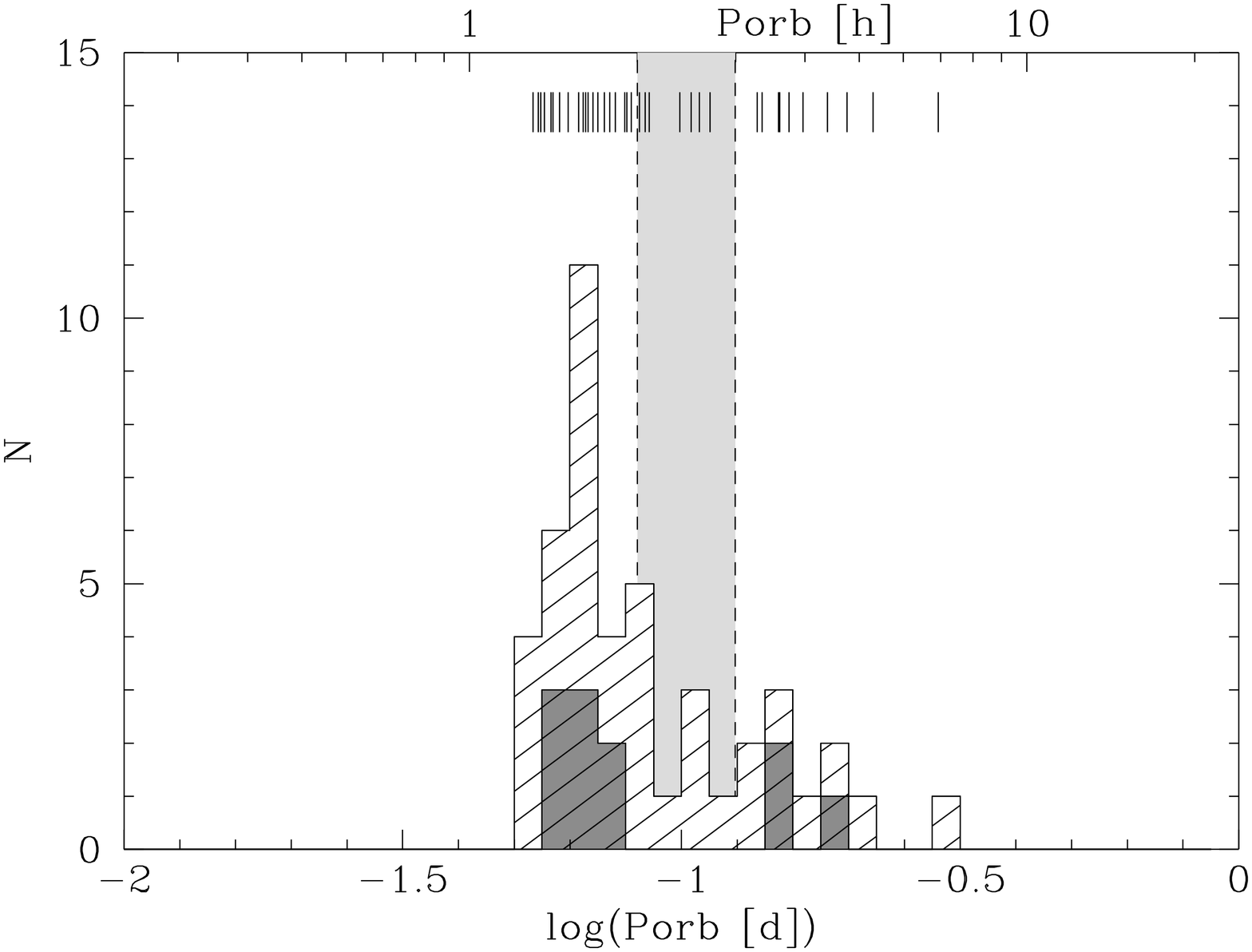}
\caption{\label{f-sdsscvs} \textit{Left:} Galactic distribution of
    CVs from the Sloan Digital Sky
    Survey. \textit{Right:} Orbital period distribution of the SDSS
    CVs. See Fig.\,\ref{f-pg} for details.}
\end{figure}

\begin{figure}
\includegraphics[angle=-90,width=0.6\columnwidth]{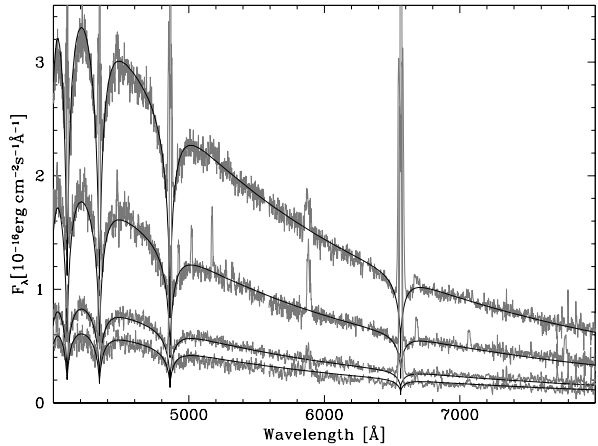}
\hspace*{-0.9cm}
\parbox[t]{0.5\columnwidth}{\caption{\label{f-sdsscvwd} 
A substantial number of the SDSS CVs are low mass transfer systems
where the white dwarf is detected in the optical spectrum. Fitting the
SDSS spectra with white dwarf models \citep{gaensickeetal95-1} results
in $T_\mathrm{eff}\simeq10\,500$\,K. Four examples are shown on the
left, from top to bottom: SDSS1238--0339, SDSS0131--0901,
SDSS0904+0355, and SDSS1610+4459.}}
\end{figure}

So far, period measurements are available only for $\sim30$\,\% of the
new SDSS CVs (Fig.\,\ref{f-sdsscvs}), and no firm conclusions should
be drawn at this point, but it appears that the period distribution of
the SDSS CV sample differs dramatically from that of the presently
known mixed bag of CVs (Fig.\,\ref{f-allcvs})~--~it is strongly
concentrated towards periods below the gap, and the gap itself is not
very pronounced. This dominant number of short-period CVs is likely to
prevail when more periods are determined, as the faint end of the SDSS
CV sample has not yet been explored, and these systems are most likely
intrinsically faint low-mass transfer systems. A substantial number of
the SDSS CVs clearly exhibit the Balmer lines of the white dwarf
accretor, and fitting those spectra with model atmospheres suggests
low effective temperatures around 11\,000\,K
(Fig.\,\ref{f-sdsscvwd}), implying very low secular mean mass transfer
rates \citep{townsley+bildsten02-2}. Whereas in CVs white dwarf
temperature determinations from optical data alone are prone to some
uncertainties (because of the contamination by the disc/hot spot), low
white dwarf temperatures are independently suggested by the fact that
several of the SDSS CVs have been found to contain pulsating ZZ\,Ceti
white dwarfs \citep[e.g.][]{woudt+warner04-1}. Low mass transfer rates
are also indicated by the absence of observed outbursts in most of the
white dwarf dominated systems. Combining all available evidence, it
appears that SDSS may be finding the missing population of old,
low-luminosity short period CVs\footnote{In fact, these low-luminosity
SDSS CVs may represent just the tip of the iceberg. The sample of
\textit{single} white dwarfs for which SDSS spectroscopy has been
obtained shows a sharp cut-off for $T_\mathrm{eff}\la10\,000$\,K
\citep{kleinmanetal04-1}, which is a clear bias in the target
selection of the SDSS spectroscopy. If the majority of CVs are
post-period minimum systems, accretion heating may not be sufficient
to keep their white dwarfs hot enough to be targeted by SDSS. It is
important to realise that while the SDSS discovered low-luminosity CVs
seem to match the predicted properties for old CVs near the minimum
period, their \textit{number} still falls significantly short with
respect to the predicted space density: none of them is within
$d\la100$\,pc.}. Establishing the properties of the complete SDSS CV
sample is of outmost importance for testing and improving our
understanding of CV evolution, but will require massive observational
effort. A first step towards this goal has been the award of the
International Time 2004/5 on La Palma for follow-up studies of SDSS
CVs (PI G\"ansicke).

\section{Other surveys of interest}
A number of additional surveys have led to the discovery of CVs, and
we can give here merely a brief and incomplete list. The First and
Second Byurakan Survey \citep{stepanianetal99-1} were objective prism
surveys similar to the HQS (Sect.\,\ref{s-hqscvs}), resulting in the
discovery of $\sim20$ CVs and CV candidates. However, almost no
detailed follow-up studies have been carried out with the exception of
the IP DW\,Cnc, independently discovered in the HQS
\citep{rodriguez-giletal04-1}. The Cal\'an-Tololo survey has been
another objective prism survey, carried out in the southern
hemisphere, leading to the identification of 16 new CVs, with the
first follow-up results just being published
\citep{tappertetal04-1}. Finally, the Hamburg-ESO survey
\citep{wisotzkietal96-1}, a southern twin of the HQS, could
potentially lead to the discovery of a substantial number of new
southern declination CVs, but so far just a single new system has been
reported  BW\,Scl\,=\,HE2350--3950 \citep{augusteijn+wisotzki97-1}.

\acknowledgements
We thank Neil Parley and Lindsey Shaw Greening, who compiled the data
used in Sect.\,\ref{s-allcvs} in an undergraduate project at the
University of Warwick; all our HQS CV survey team members, and Paula
Szkody for numerous discussions and communications on SDSS CVs. This
research has been supported by a PPARC Advanced Fellowship.

\bibliographystyle{aa}

\end{document}